\documentclass[journal]{IEEEtran}
\usepackage{amssymb}
\usepackage{latexsym}
\usepackage{amsmath}
\usepackage{empheq}
\usepackage{subfig,caption,graphicx,color,epsfig,graphicx}

\usepackage{url}
\usepackage{xcolor}
\definecolor{newcolor}{rgb}{.8,.349,.1}

\newtheorem{theorem}{Theorem}[section]
\newtheorem{corollary}{Corollary}[theorem]
\newtheorem{lemma}[theorem]{Lemma}

\usepackage{algorithmic}
\begin{document}
\title{2D Sinusoidal Parameter Estimation with Offset Term}

\author{A.~Pasha~Hosseinbor
        and~Renat~Zhdanov
\thanks{A. P. Hosseinbor and R. Zhdanov are with Bio-Key International Inc., Eagan, MN, USA}}

\maketitle

\begin{abstract}
We consider the parameter estimation of a 2D sinusoid. Although sinusoidal parameter estimation has been extensively studied, our model differs from those examined in the available literature by the inclusion of an offset term. We derive both the maximum likelihood estimation (MLE) solution and the Cramer-Rao lower bound (CRLB) on the variance of the model's estimators.
\end{abstract}


\IEEEpeerreviewmaketitle

\section{Introduction}

In this paper, we examine the problem of parameter estimation of a 2D sinusoid. Although sinusoidal parameter estimation has been extensively studied \cite{rife.1974,rife.1976,lang.1980,stoica.1989,hainsworth.2003}, our model differs slightly from those examined in the available literature by the inclusion of an offset term. We derive both the MLE solution and the Cramer-Rao lower bound (CRLB) on the variance our model's estimators, and then implement our approach on several fingerprint images of varying quality.  

We specifically consider the discrete 2D sinusoidal signal 

\begin{equation}
\label{eq:model}
f(x,y)=A\sin(2\pi(f_{0}x+f_{1}y)+\phi)+B,
\end{equation}
where $x=0,...,N-1$, $y=0,...,N-1$; $\boldsymbol{\theta}=(A\;B\;\phi\;f_{0}\;f_{1})$ is the vector of parameters to be estimated: $A$ is the amplitude of the sinusoid, $B$ is its offset, $\phi$ is its phase shift, and ${\bf f}=(f_{1}\;\;f_{0})^{T}$ is its frequency. Such a model could describe the signal intensity at pixel $(x,y)$ of an $N \times N$ image. The main difference between Eq. (\ref{eq:model}) and those studied in \cite{rife.1974,rife.1976,lang.1980,stoica.1989,hainsworth.2003} is the inclusion of the offset term $B$.

Eq. (\ref{eq:model}) arises in fingerprint biometrics. Fingerprint texture is characterized by the periodic flow of ridges and furrows, so it contains both frequency and orientation information; the frequency content is due to the inter-ridge spacing present in the fingerprint, while the orientation is due to the flow pattern exhibited by the ridges. If an acquired (gray-level) 2D fingerprint image is partitioned into sub-blocks, where each sub-block contains a ridge segment, the gray level intensity variations can be modeled via Eq. (\ref{eq:model}), whose parameters characterize the enclosed ridge's frequency and orientation within the sub-block. 

\section{Theory}

The following theorems will prove useful in our derivations of both the CRLB and MLE of $\boldsymbol{\theta}$.

%\footnotesize
%\begin{lemma}
%For $\omega \in [0,2\pi]$,
%\[
%\frac{1}{N}\sum_{n=0}^{N-1}\cos(\omega n+\phi)=\frac{\sin(\omega/2-\phi)+\sin(\omega(N-1/2)+\phi)}{2N\sin(\omega/2)} 
%\]
%
%\[
%\frac{1}{N}\sum_{n=0}^{N-1}\sin(\omega n+\phi) = \frac{\cos(\omega/2-\phi)-\cos(\omega(N-1/2)+\phi)}{2N\sin(\omega/2)} 
%\]
%\end{lemma}

%\begin{corollary}
%\text{For} $f \in [0,1]$ \text{and} $k\geq0$,
%\[
%  \lim_{N\to\infty}\frac{1}{N^{k+1}}\sum_{n=0}^{N-1}n^{k}\cos(2\pi fn+\phi)= \left\{\def\arraystretch{1.2}%
%  \begin{array}{@{}c@{\quad}l@{}}
%    \frac{1}{k+1}\cos(\phi) & \text{$f=0,1$}\\
%    0 & \text{$f \neq 0,1$}\\
%  \end{array}\right.
%\]
%
%\[
%  \lim_{N\to\infty}\frac{1}{N^{k+1}}\sum_{n=0}^{N-1}n^{k}\sin(2\pi fn+\phi)= \left\{\def\arraystretch{1.2}%
%  \begin{array}{@{}c@{\quad}l@{}}
%    \frac{1}{k+1}\sin(\phi) & \text{$f=0,1$}\\
%    0 & \text{$f \neq 0,1$}\\
%  \end{array}\right.
%\]
%\end{corollary}
%\normalsize 
%In the course of our derivations of the CRLB and MLE, we will encounter trigonometric sums of the form $\sum_{n}\text{cis}(2\pi fn+\phi)$ and $\sum_{n}\text{cis}(4\pi fn +2\phi)$. For the former, if $f=0$ or $f=1$, the sum will be non-zero. For the latter, if $f=0$, $f=1/2$, or $f=1$, the sum will be non-zero. Hence, we require that $f_{0},f_{1} \neq 0,1/2,1$.

\footnotesize
\begin{lemma}
For $\omega \in [0,2\pi]$,
\[
\frac{1}{N}\sum_{n=0}^{N-1}e^{i(\omega n+\phi)}=\frac{e^{i(\pi/2 - \omega/2  + \phi)}+e^{-i(\pi/2 - \omega(N-1/2) - \phi)}}{2N\sin(\omega/2)} 
\]
\end{lemma}

\begin{corollary}
\text{For} $f \in [0,1]$ \text{and integer} $k\geq0$,
\[
  \lim_{N\to\infty}\frac{1}{N^{k+1}}\sum_{n=0}^{N-1}n^{k}e^{i(2\pi fn+\phi)}= \left\{\def\arraystretch{1.2}%
  \begin{array}{@{}c@{\quad}l@{}}
    \frac{1}{k+1}e^{i\phi} & \text{$f=0,1$}\\
    0 & \text{$f \neq 0,1$}\\
  \end{array}\right.
\]
\end{corollary}
\normalsize

\subsection{Crammer-Rao Lower Bound (CRLB) of Estimator $\boldsymbol{\theta}$}

Consider the $p \times 1$ vector parameter $\boldsymbol{\theta}=(\theta_{1}\;\dots\;\theta_{p})$. We will assume that the estimator $\boldsymbol{\hat{\theta}}$ is unbiased. The CRLB gives a lower bound on the variance of any unbiased estimator, and the CRLB of estimator $\hat{\theta_{i}}$ is 

\begin{equation}
\label{eq:crlb}
\text{var}(\hat{\theta_{i}}) \geq [\boldsymbol{\eta}^{-1}(\boldsymbol{\theta})]_{ii},
\end{equation}
where $\boldsymbol{\eta(\theta)}$ is the $p$ x $p$ Fisher information matrix; it is defined as

\begin{equation}
\label{eq:FI}
[\boldsymbol{\eta}(\boldsymbol{\theta})]_{ij}=-E \left[ \frac{\partial^{2}\ln p({\bf x};\boldsymbol{\theta})}{\partial\theta_{i}\partial\theta_{j}} \right]
\end{equation}
for $i=1,2,\dots,p$ and $j=1,2,\dots,p$. 

We consider the signal 

\small
\begin{equation}
\label{eq:signal}
s(x,y)=f(x,y)+w(x,y), \;\;\;\;\; x=0,\dots,N-1; \; y=0,\dots,N-1
\end{equation}
\normalsize
where $f(x,y)$ is given by Eq. (\ref{eq:model}) and $w(x,y)$ is the noise. Since we assume the noise is white Gaussian, i.e. $w(x,y)=\frac{1}{2\pi\sigma^{2}}\exp(-\frac{x^{2}+y^{2}}{2\sigma^{2}})$, we have $s(x,y) \sim \mathcal{N}(f(x,y),\sigma^{2})$. 

Denote ${\bf z}=\text{vec} \{f(x,y); \; x=0,\dots,N-1, \; y=0,\dots,N-1\}$ and ${\bf w}=\text{vec} \{w(x,y); \; x=0,\dots,N-1, \; y=0,\dots,N-1\}$; both are of dimension $N^{2}$ x $1$. Then Eq. (\ref{eq:signal}) can be rewritten in vector form as ${\bf s}={\bf z}+{\bf w}$, where the signal measurements ${\bf s} \sim \mathcal{N}_{N^{2}}({\bf z}(\boldsymbol{\theta}),\sigma^{2}{\bf I}_{N^{2} \times N^{2}})$. Then the log-likelihood function of $\boldsymbol{\theta}$ (ignoring the fixed term) is

%\scriptsize
%\begin{align*}
%\ln p({\bf s};\boldsymbol{\theta})= {} & \ln(c)-\frac{1}{2\sigma^{2}}({\bf s}-{\bf z}(\boldsymbol{\theta}))^{T}({\bf s}-{\bf z}(\boldsymbol{\theta})) \nonumber \\ = {} & -\frac{1}{2\sigma^{2}}\sum_{x=0}^{N-1}\sum_{y=0}^{N-1} ( s^{2}(x,y)-2s(x,y)A\sin(2\pi(f_{0}x+f_{1}y)+\phi) \\ & -2s(x,y)B+A^{2}\sin^{2}(2\pi(f_{0}x+f_{1}y)+\phi) \\ & +2AB\sin(2\pi(f_{0}x+f_{1}y)+\phi)+B^{2} )
%\end{align*}
%\normalsize
\scriptsize
\begin{align*}
\ln p({\bf s};\boldsymbol{\theta})= {} & -\frac{1}{2\sigma^{2}}\sum_{x=0}^{N-1}\sum_{y=0}^{N-1} ( s^{2}(x,y)-2s(x,y)A\sin(2\pi(f_{0}x+f_{1}y)+\phi) \\ & -2s(x,y)B+A^{2}\sin^{2}(2\pi(f_{0}x+f_{1}y)+\phi) \\ & +2AB\sin(2\pi(f_{0}x+f_{1}y)+\phi)+B^{2} )
\end{align*}
\normalsize
%where we have dropped the fixed term, $\ln(c)$. 

We now derive the elements forming the Fisher information matrix, given by Eq. (\ref{eq:FI}).
\newline

\begin{enumerate}

\item $E \left[ \frac{\partial^{2}\ln p({\bf s};\boldsymbol{\theta})}{\partial A^{2}} \right]$:
\footnotesize
\begin{align*}
\frac{\partial^{2}\ln p({\bf s};\boldsymbol{\theta})}{\partial A^{2}} = {} & -\frac{1}{\sigma^{2}}\sum_{x=0}^{N-1}\sum_{y=0}^{N-1}\left( \frac{1}{2}-\frac{1}{2}\cos(4\pi(f_{0}x+f_{1}y)+2\phi) \right)  \\ = {} & -\frac{N^{2}}{2\sigma^{2}}+\frac{1}{2\sigma^{2}}\sum_{x}\cos(4\pi f_{0}x+2\phi)\sum_{y}\cos(4\pi f_{1}y) \\ & -\frac{1}{2\sigma^{2}}\sum_{x}\sin(4\pi f_{0}x+2\phi)\sum_{y}\sin(4\pi f_{1}y) \\ \approx {} & -\frac{N^{2}}{2\sigma^{2}},
\end{align*}
\normalsize
where we have used the approximation that $\frac{1}{N}\sum_{x=0}^{N-1}\sin(4\pi f_{0}x+2\phi) \approx 0$ for large $N$ and $f_{0} \neq 0,1/2, 1$. 

\begin{align*}
\boxed{ E \left[ \frac{\partial^{2}\ln p({\bf s};\boldsymbol{\theta})}{\partial A^{2}} \right] \approx -\frac{N^{2}}{2\sigma^{2}}}
\end{align*}

\item $E \left[ \frac{\partial^{2}\ln p({\bf s};\boldsymbol{\theta})}{\partial A\partial B} \right]$:
\footnotesize
\begin{align*}
\frac{\partial^{2}\ln p({\bf s};\boldsymbol{\theta})}{\partial A\partial B} &= -\frac{1}{\sigma^{2}}\sum_{x=0}^{N-1}\sum_{y=0}^{N-1}\sin(2\pi(f_{0}x+f_{1}y)+\phi) \approx 0,
\end{align*}
\normalsize
where we have employed the approximation that $\frac{1}{N}\sum_{x=0}^{N-1}\sin(2\pi f_{0}x+\phi) \approx 0$ for large $N$ and $f_{0} \neq 0,1$. 

\begin{align*}
\boxed{ E \left[ \frac{\partial^{2}\ln p({\bf s};\boldsymbol{\theta})}{\partial A\partial B} \right] \approx 0}
\end{align*}

\item $E \left[ \frac{\partial^{2}\ln p({\bf s};\boldsymbol{\theta})}{\partial A \partial\phi}\right]$:
\small
\begin{align*}
\frac{\partial^{2}\ln p({\bf s};\boldsymbol{\theta})}{\partial A \partial\phi} = {} & -\frac{1}{2\sigma^{2}}\sum_{x=0}^{N-1}\sum_{y=0}^{N-1}(2A\sin(4\pi(f_{0}x+f_{1}y)+2\phi) \\ & +2\cos(2\pi(f_{0}x+f_{1}y)+\phi)(B-s(x,y))) 
\end{align*}
\normalsize

\scriptsize
\begin{align*}
E \left[ \frac{\partial^{2}\ln p({\bf s};\boldsymbol{\theta})}{\partial A \partial\phi}\right] = {} & -\frac{1}{2\sigma^{2}}\sum_{x=0}^{N-1}\sum_{y=0}^{N-1}(2A\sin(4\pi(f_{0}x+f_{1}y)+2\phi) \\ & +2\cos(2\pi(f_{0}x+f_{1}y)+\phi)(B-E[s(x,y)])) \\ = {} & -\frac{A}{2\sigma^{2}}\sum_{x}\sum_{y}\sin(4\pi(f_{0}x+f_{1}y)+2\phi) \\ \approx {} & \boxed{0}
\end{align*} 
\normalsize

\item $E \left[ \frac{\partial^{2}\ln p({\bf s};\boldsymbol{\theta})}{\partial A \partial f_{0}} \right]$:
\footnotesize
\begin{align*}
\frac{\partial^{2}\ln p({\bf s};\boldsymbol{\theta})}{\partial A \partial f_{0}} = {} & -\frac{1}{2\sigma^{2}}\sum_{x=0}^{N-1}\sum_{y=0}^{N-1}(4\pi Ax\sin(4\pi(f_{0}x+f_{1}y)+2\phi) \\ & +4\pi x\cos(2\pi(f_{0}x+f_{1}y)+\phi)(B-s(x,y)))
\end{align*}
\normalsize

\footnotesize
\begin{align*}
E \left[ \frac{\partial^{2}\ln p({\bf s};\boldsymbol{\theta})}{\partial A \partial f_{0}} \right] = {} & -\frac{1}{2\sigma^{2}}\sum_{x,y}(4\pi Ax\sin(4\pi(f_{0}x+f_{1}y)+2\phi) \\ & +4\pi x\cos(2\pi(f_{0}x+f_{1}y)+\phi)(B-E[s(x,y)])) \\ = {} & -\frac{A\pi}{\sigma^{2}}\sum_{x=0}^{N-1}\sum_{y=0}^{N-1}x\sin(4\pi(f_{0}x+f_{1}y)+2\phi) \\ \approx {} & \boxed{0},
\end{align*}
\normalsize
where we have used the approximation that $\frac{1}{N^{2}}\sum_{x=0}^{N-1}x\sin(4\pi f_{0}x+2\phi) \approx 0$ for large $N$ and $f_{0} \neq 0,1/2$.
\newline 

\item $E \left[ \frac{\partial^{2}\ln p({\bf s};\boldsymbol{\theta})}{\partial A \partial f_{1}} \right] \approx \boxed{0}$ (calculation similar to (4)) 
%\footnotesize
%\begin{align*}
%\frac{\partial^{2}\ln p({\bf s};\boldsymbol{\theta})}{\partial A \partial f_{1}} = {} & -\frac{1}{2\sigma^{2}}\sum_{x=0}^{N-1}\sum_{y=0}^{N-1}(4\pi Ay\sin(4\pi(f_{0}x+f_{1}y)+2\phi) \\ & +4\pi y\cos(2\pi(f_{0}x+f_{1}y)+\phi)(B-s(x,y)))
%\end{align*}
%\normalsize 
%
%\footnotesize
%\begin{align*}
%E \left[ \frac{\partial^{2}\ln p({\bf s};\boldsymbol{\theta})}{\partial A \partial f_{1}} \right] = {} &-\frac{1}{2\sigma^{2}}\sum_{x=0}^{N-1}\sum_{y=0}^{N-1}(4\pi Ay\sin(4\pi(f_{0}x+f_{1}y)+2\phi) \\ & +4\pi y\cos(2\pi(f_{0}x+f_{1}y)+\phi)(B-E[s(x,y)])) \\ = {} & -\frac{A\pi}{\sigma^{2}}\sum_{x}\sum_{y}y\sin(4\pi(f_{0}x+f_{1}y)+2\phi) \\ \approx {} & \boxed{0}
%\end{align*}
%\normalsize

%\begin{align*}
%E \left[ \frac{\partial^{2}\ln p({\bf s};\boldsymbol{\theta})}{\partial A \partial f_{1}} \right] \approx {} & \boxed{0}
%\end{align*}

\item $E \left[ \frac{\partial^{2}\ln p({\bf s};\boldsymbol{\theta})}{\partial B^{2}} \right] = \boxed{-\frac{N^{2}}{\sigma^{2}}}$
%\begin{align*}
%\frac{\partial^{2}\ln p({\bf s};\boldsymbol{\theta})}{\partial B^{2}} = -\frac{N^{2}}{\sigma^{2}}
%\end{align*}
%
%\begin{align*}
%\boxed{E \left[ \frac{\partial^{2}\ln p({\bf s};\boldsymbol{\theta})}{\partial B^{2}} \right] = -\frac{N^{2}}{\sigma^{2}}}
%\end{align*}

\item $E \left[ \frac{\partial^{2}\ln p({\bf s};\boldsymbol{\theta})}{\partial B\partial \phi} \right] \approx \boxed{0}$ (calculation similar to (2))

%\footnotesize
%\begin{align*}
%\frac{\partial^{2}\ln p({\bf s};\boldsymbol{\theta})}{\partial B\partial \phi} = -\frac{1}{\sigma^{2}}\sum_{x=0}^{N-1}\sum_{y=0}^{N-1}\cos(2\pi(f_{0}x+f_{1}y)+\phi) \approx 0
%\end{align*}
%\normalsize

%\begin{align*}
%\boxed{E \left[ \frac{\partial^{2}\ln p({\bf s};\boldsymbol{\theta})}{\partial B\partial \phi} \right] \approx 0}
%\end{align*}

\item $E \left[ \frac{\partial^{2}\ln p({\bf s};\boldsymbol{\theta})}{\partial B\partial f_{0}} \right] \approx \boxed{0}$ (calculation similar to (4))
%\footnotesize
%\begin{align*}
%\frac{\partial^{2}\ln p({\bf s};\boldsymbol{\theta})}{\partial B\partial f_{0}} = -\frac{2\pi A}{\sigma^{2}}\sum_{x=0}^{N-1}\sum_{y=0}^{N-1}x\cos(2\pi(f_{0}x+f_{1}y)+\phi)
%\end{align*}
%\normalsize
%
%\begin{align*}
%\boxed{E \left[ \frac{\partial^{2}\ln p({\bf s};\boldsymbol{\theta})}{\partial B\partial f_{0}} \right] \approx 0}
%\end{align*}

\item $E \left[ \frac{\partial^{2}\ln p({\bf s};\boldsymbol{\theta})}{\partial B\partial f_{1}} \right] \approx \boxed{0}$ (calculation similar to (4))
%\footnotesize
%\begin{align*}
%\frac{\partial^{2}\ln p({\bf s};\boldsymbol{\theta})}{\partial B\partial f_{1}} = -\frac{2\pi A}{\sigma^{2}}\sum_{x=0}^{N-1}\sum_{y=0}^{N-1}y\cos(2\pi(f_{0}x+f_{1}y)+\phi)
%\end{align*}
%\normalsize

%\begin{align*}
%\boxed{E \left[ \frac{\partial^{2}\ln p({\bf s};\boldsymbol{\theta})}{\partial B\partial f_{1}} \right] \approx 0}
%\end{align*}

\item $E \left[ \frac{\partial^{2}\ln p({\bf s};\boldsymbol{\theta})}{\partial \phi^{2}} \right]$:
\footnotesize
\begin{align*}
\frac{\partial^{2}\ln p({\bf s};\boldsymbol{\theta})}{\partial \phi^{2}} = {} & -\frac{1}{2\sigma^{2}}\sum_{x=0}^{N-1}\sum_{y=0}^{N-1}(2A^{2}\cos(4\pi(f_{0}x+f_{1}y)+2\phi) \\ & +2A\sin(2\pi(f_{0}x+f_{1}y)+\phi)(s(x,y)-B))
\end{align*}
\normalsize

%\scriptsize
%\begin{align*}
%E \left[ \frac{\partial^{2}\ln p({\bf s};\boldsymbol{\theta})}{\partial \phi^{2}} \right] = {} & -\frac{1}{2\sigma^{2}}\sum_{x=0}^{N-1}\sum_{y=0}^{N-1}(2A^{2}\cos(4\pi(f_{0}x+f_{1}y)+2\phi) \\ & +2A\sin(2\pi(f_{0}x+f_{1}y)+\phi)(E[s(x,y)]-B)) \\ = {} & -\frac{1}{2\sigma^{2}}\sum_{x=0}^{N-1}\sum_{y=0}^{N-1}(2A^{2}\cos(4\pi(f_{0}x+f_{1}y)+2\phi) \\ & +2A^{2}\sin^{2}(2\pi(f_{0}x+f_{1}y)+\phi)) \\ = {} & -\frac{1}{2\sigma^{2}}\sum_{x=0}^{N-1}\sum_{y=0}^{N-1}(2A^{2}\cos^{2}(2\pi(f_{0}x+f_{1}y)+\phi) \\ = {} &  -\frac{A^{2}}{2\sigma^{2}}\sum_{x=0}^{N-1}\sum_{y=0}^{N-1}(1+\cos(4\pi(f_{0}x+f_{1}y)+2\phi)) \\ \approx {} &  \boxed{-\frac{A^{2}N^{2}}{2\sigma^{2}}},
%\end{align*}
%\normalsize
\scriptsize
\begin{align*}
E \left[ \frac{\partial^{2}\ln p({\bf s};\boldsymbol{\theta})}{\partial \phi^{2}} \right] = {} & -\frac{1}{2\sigma^{2}}\sum_{x=0}^{N-1}\sum_{y=0}^{N-1}(2A^{2}\cos(4\pi(f_{0}x+f_{1}y)+2\phi) \\ & +2A\sin(2\pi(f_{0}x+f_{1}y)+\phi)(E[s(x,y)]-B)) \\ = {} &  -\frac{A^{2}}{2\sigma^{2}}\sum_{x=0}^{N-1}\sum_{y=0}^{N-1}(1+\cos(4\pi(f_{0}x+f_{1}y)+2\phi)) \\ \approx {} &  \boxed{-\frac{A^{2}N^{2}}{2\sigma^{2}}},
\end{align*}
\normalsize
where we have employed the identity $\cos^{2}(x)=\frac{1+\cos(2x)}{2}$.
\newline

\item $E \left[ \frac{\partial^{2}\ln p({\bf s};\boldsymbol{\theta})}{\partial \phi \partial f_{0}} \right]$:
\footnotesize
\begin{align*}
\frac{\partial^{2}\ln p({\bf s};\boldsymbol{\theta})}{\partial \phi \partial f_{0}} = {} & -\frac{1}{2\sigma^{2}}\sum_{x=0}^{N-1}\sum_{y=0}^{N-1}(4\pi A^{2}x\cos(4\pi(f_{0}x+f_{1}y)+2\phi) \\ & +4\pi Ax\sin(2\pi(f_{0}x+f_{1}y)+\phi)(s(x,y)-B))
\end{align*}
\normalsize

%\scriptsize
%\begin{align*}
%E \left[ \frac{\partial^{2}\ln p({\bf s};\boldsymbol{\theta})}{\partial \phi \partial f_{0}} \right]= {} & -\frac{1}{2\sigma^{2}}\sum_{x=0}^{N-1}\sum_{y=0}^{N-1}(4\pi A^{2}x\cos(4\pi(f_{0}x+f_{1}y)+2\phi) \\ & +4\pi Ax\sin(2\pi(f_{0}x+f_{1}y)+\phi)(E[s(x,y)]-B)) \\ = {} & -\frac{1}{2\sigma^{2}}\sum_{x=0}^{N-1}\sum_{y=0}^{N-1}(4\pi A^{2}x\cos(4\pi(f_{0}x+f_{1}y)+2\phi) \\  &+4\pi A^{2}x\sin^{2}(2\pi(f_{0}x+f_{1}y)+\phi)) \\ = {} & -\frac{1}{2\sigma^{2}}\sum_{x=0}^{N-1}\sum_{y=0}^{N-1}4\pi A^{2}x\cos^{2}(2\pi(f_{0}x+f_{1}y)+\phi) \\ = {} & -\frac{\pi A^{2}}{\sigma^{2}}\sum_{x=0}^{N-1}\sum_{y=0}^{N-1}(x+\cos(4\pi(f_{0}x+f_{1}y)+2\phi)) \\ \approx {} & -\frac{\pi A^{2}}{\sigma^{2}}\sum_{x=0}^{N-1}\sum_{y=0}^{N-1}x =  \boxed{-\frac{\pi A^{2}N^{2}(N-1)}{2\sigma^{2}}}
%\end{align*}
%\normalsize

\scriptsize
\begin{align*}
E \left[ \frac{\partial^{2}\ln p({\bf s};\boldsymbol{\theta})}{\partial \phi \partial f_{0}} \right]= {} & -\frac{1}{2\sigma^{2}}\sum_{x=0}^{N-1}\sum_{y=0}^{N-1}(4\pi A^{2}x\cos(4\pi(f_{0}x+f_{1}y)+2\phi) \\ & +4\pi Ax\sin(2\pi(f_{0}x+f_{1}y)+\phi)(E[s(x,y)]-B)) \\ = {} & -\frac{\pi A^{2}}{\sigma^{2}}\sum_{x=0}^{N-1}\sum_{y=0}^{N-1}(x+\cos(4\pi(f_{0}x+f_{1}y)+2\phi)) \\ \approx {} & -\frac{\pi A^{2}}{\sigma^{2}}\sum_{x=0}^{N-1}\sum_{y=0}^{N-1}x =  \boxed{-\frac{\pi A^{2}N^{2}(N-1)}{2\sigma^{2}}}
\end{align*}
\normalsize

\item $E \left[ \frac{\partial^{2}\ln p({\bf s};\boldsymbol{\theta})}{\partial \phi \partial f_{1}} \right] \approx \boxed{-\frac{\pi A^{2}N^{2}(N-1)}{2\sigma^{2}}}$ (calculation similar to (11))
%\scriptsize
%\begin{align*}
%\frac{\partial^{2}\ln p({\bf s};\boldsymbol{\theta})}{\partial \phi \partial f_{1}} = {} & -\frac{1}{2\sigma^{2}}\sum_{x=0}^{N-1}\sum_{y=0}^{N-1}(4\pi A^{2}y\cos(4\pi(f_{0}x+f_{1}y)+2\phi) \\ & +4\pi Ay\sin(2\pi(f_{0}x+f_{1}y)+\phi)(s(x,y)-B))
%\end{align*}
%\normalsize

%\begin{align*}
%\boxed{E \left[ \frac{\partial^{2}\ln p({\bf s};\boldsymbol{\theta})}{\partial \phi \partial f_{1}} \right] \approx -\frac{\pi A^{2}N^{2}(N-1)}{2\sigma^{2}}}
%\end{align*}

\item $E \left[ \frac{\partial^{2}\ln p({\bf s};\boldsymbol{\theta})}{\partial f_{0}^{2}} \right]$:
\scriptsize
\begin{align*}
\frac{\partial^{2}\ln p({\bf s};\boldsymbol{\theta})}{\partial f_{0}^{2}} = {} & -\frac{1}{2\sigma^{2}}\sum_{x=0}^{N-1}\sum_{y=0}^{N-1}(8\pi^{2}A^{2}x^{2}\cos(4\pi(f_{0}x+f_{1}y)+2\phi) \\ & +8\pi^{2} Ax^{2}\sin(2\pi(f_{0}x+f_{1}y)+\phi)(s(x,y)-B))
\end{align*}
\normalsize

%\scriptsize
%\begin{align*}
%E \left[ \frac{\partial^{2}\ln p({\bf s};\boldsymbol{\theta})}{\partial f_{0}^{2}} \right] = {} & -\frac{1}{2\sigma^{2}}\sum_{x,y}(8\pi^{2}A^{2}x^{2}\cos(4\pi(f_{0}x+f_{1}y)+2\phi) \\ & +8\pi^{2} Ax^{2}\sin(2\pi(f_{0}x+f_{1}y)+\phi)(E[s(x,y)] \\ {}  & -B)) \\ = {} & -\frac{1}{2\sigma^{2}}\sum_{x,y}8\pi^{2}A^{2}x^{2}\cos^{2}(2\pi(f_{0}x+f_{1}y)+\phi) 
%\\ = {} & -\frac{2\pi^{2}A^{2}}{\sigma^{2}}\sum_{x,y}(x^{2}+x^{2}\cos(4\pi(f_{0}x+f_{1}y)+2\phi)) \\ \approx {} & -\frac{2\pi^{2}A^{2}}{\sigma^{2}}\sum_{x=0}^{N-1}\sum_{y=0}^{N-1}x^{2} \\ = {} & \boxed{-\frac{\pi^{2}A^{2}N^{2}(N-1)(2N-1)}{3\sigma^{2}}},
%\end{align*}
%\normalsize
\scriptsize
\begin{align*}
E \left[ \frac{\partial^{2}\ln p({\bf s};\boldsymbol{\theta})}{\partial f_{0}^{2}} \right] = {} & -\frac{1}{2\sigma^{2}}\sum_{x,y}(8\pi^{2}A^{2}x^{2}\cos(4\pi(f_{0}x+f_{1}y)+2\phi) \\ & +8\pi^{2} Ax^{2}\sin(2\pi(f_{0}x+f_{1}y)+\phi)(E[s(x,y)] \\ {}  & -B)) \\ = {} & -\frac{2\pi^{2}A^{2}}{\sigma^{2}}\sum_{x,y}(x^{2}+x^{2}\cos(4\pi(f_{0}x+f_{1}y)+2\phi)) \\ \approx {} & -\frac{2\pi^{2}A^{2}}{\sigma^{2}}\sum_{x=0}^{N-1}\sum_{y=0}^{N-1}x^{2} \\ = {} & \boxed{-\frac{\pi^{2}A^{2}N^{2}(N-1)(2N-1)}{3\sigma^{2}}},
\end{align*}
\normalsize
where we have used the approximation that $\frac{1}{N^{3}}\sum_{x=0}^{N-1}x^{2}\cos(4\pi f_{0}x+2\phi) \approx 0$ for large $N$ and $f_{0} \neq 0,1/2$
\newline

\item $E \left[ \frac{\partial^{2}\ln p({\bf s};\boldsymbol{\theta})}{\partial f_{0}\partial f_{1}} \right]$:
\scriptsize
\begin{align*}
\frac{\partial^{2}\ln p({\bf s};\boldsymbol{\theta})}{\partial f_{0} \partial f_{1}} = {} & -\frac{1}{2\sigma^{2}}\sum_{x=0}^{N-1}\sum_{y=0}^{N-1}(8\pi^{2}A^{2}xy\cos(4\pi(f_{0}x+f_{1}y)+2\phi) \\ &+8\pi^{2} Axy\sin(2\pi(f_{0}x+f_{1}y)+\phi)(s(x,y)-B))
\end{align*}
\normalsize

\scriptsize 
\begin{align*}
E \left[ \frac{\partial^{2}\ln p({\bf s};\boldsymbol{\theta})}{\partial f_{0}\partial f_{1}} \right] = {} & -\frac{2\pi^{2}A^{2}}{\sigma^{2}}\sum_{x=0}^{N-1}\sum_{y=0}^{N-1}(xy+xy\cos(4\pi(f_{0}x+f_{1}y) \\ +2\phi)) \approx {} & -\frac{2\pi^{2}A^{2}}{\sigma^{2}}\sum_{x=0}^{N-1}\sum_{y=0}^{N-1}xy \\ = {} & \boxed{-\frac{\pi^{2}A^{2}N^{2}(N-1)^{2}}{2\sigma^{2}}}
\end{align*}
\normalsize 

\item $E \left[ \frac{\partial^{2}\ln p({\bf s};\boldsymbol{\theta})}{\partial f_{1}^{2}} \right] \approx \boxed{-\frac{\pi^{2}A^{2}N^{2}(N-1)(2N-1)}{3\sigma^{2}}}$ (calculation similar to (13))
%\scriptsize
%\begin{align*}
%\frac{\partial^{2}\ln p({\bf s};\boldsymbol{\theta})}{\partial f_{1}^{2}} = {} & -\frac{1}{2\sigma^{2}}\sum_{x=0}^{N-1}\sum_{y=0}^{N-1}(8\pi^{2}A^{2}y^{2}\cos(4\pi(f_{0}x+f_{1}y)+2\phi) \\ & +8\pi^{2} Ay^{2}\sin(2\pi(f_{0}x+f_{1}y)+\phi)(s(x,y)-B))
%\end{align*}
%\normalsize

%\begin{align*}
%\boxed{E \left[ \frac{\partial^{2}\ln p({\bf s};\boldsymbol{\theta})}{\partial f_{1}^{2}} \right] \approx -\frac{\pi^{2}A^{2}N^{2}(N-1)(2N-1)}{3\sigma^{2}}}
%\end{align*}

\end{enumerate}

%Hence, the Fisher information matrix is
%
%\scriptsize
%$$%\[
% \boldsymbol{\eta}(\boldsymbol{\theta})=\frac{N^{2}}{\sigma^{2}}
%\begin{pmatrix}
%\frac{1}{2} & 0 & 0 & 0 & 0 \\ 
%0 & 1 & 0 & 0 & 0 \\
%0 & 0 & \frac{A^{2}}{2} & \frac{\pi A^{2}(N-1)}{2} & \frac{\pi A^{2}(N-1)}{2} \\
%0 & 0 & \frac{\pi A^{2}(N-1)}{2} & \frac{\pi^{2} A^{2}(N-1)(2N-1)}{3} & \frac{\pi^{2} A^{2}(N-1)^{2}}{2} \\
%0 & 0 & \frac{\pi A^{2}(N-1)}{2} & \frac{\pi^{2} A^{2}(N-1)^{2}}{2} & \frac{\pi^{2} A^{2}(N-1)(2N-1)}{3} 
%\end{pmatrix}
%%\]
%$$
%\normalsize
Noting that the determinant is $| \boldsymbol{\eta}(\boldsymbol{\theta}) | = \frac{\pi^{4}A^{6}N^{10}(N^{2}-1)^{2}}{144\sigma^{10}}$, matrix inversion yields 

\footnotesize
$$%\[
 \boldsymbol{\eta}^{-1}(\boldsymbol{\theta})=\frac{\sigma^{2}}{N^{2}}
\begin{pmatrix}
2& 0 & 0 & 0 & 0 \\ 
0 & 1 & 0 & 0 & 0 \\
0 & 0 & \frac{2(7N-5)}{A^{2}(N+1)} & \frac{-6}{\pi A^{2}(N+1)} & \frac{-6}{\pi A^{2}(N+1)} \\
0 & 0 & \frac{-6}{\pi A^{2}(N+1)} & \frac{6}{\pi^{2} A^{2}(N^{2}-1)} & 0 \\
0 & 0 & \frac{-6}{\pi A^{2}(N+1)} & 0 &  \frac{6}{\pi^{2} A^{2}(N^{2}-1)}
\end{pmatrix}.
%\]
$$
\normalsize
Hence, the CRLB of our estimator $\boldsymbol{\hat{\theta}}$ in Eq. (\ref{eq:model}), under the assumption of white Gaussian noise $\mathcal{N}(0,\sigma^{2})$, is

%\begin{equation}
%\begin{split}
%\text{var}(\hat{A}) \geq {} & \frac{2\sigma^{2}}{N^{2}}  \\
%\text{var}(\hat{B}) \geq {} & \frac{\sigma^{2}}{N^{2}}  \\
%\text{var}(\hat{\phi}) \geq {} & \frac{2(7N-5)\sigma^{2}}{A^{2}N^{2}(N+1)} \\
%\text{var}(\hat{f_{0}}) \geq {} & \frac{6\sigma^{2}}{\pi^{2} A^{2}N^{2}(N^{2}-1)} \\
%\text{var}(\hat{f_{1}}) \geq {} & \frac{6\sigma^{2}}{\pi^{2} A^{2}N^{2}(N^{2}-1)}
%\end{split}
%\end{equation}
\begin{empheq}[box=\fbox]{align}
\text{var}(\hat{A}) &\geq \frac{2\sigma^{2}}{N^{2}} \nonumber \\
\text{var}(\hat{B}) &\geq \frac{\sigma^{2}}{N^{2}}  \nonumber \\
\text{var}(\hat{\phi}) &\geq \frac{2(7N-5)\sigma^{2}}{A^{2}N^{2}(N+1)} \nonumber \\
\text{var}(\hat{f_{0}}) &\geq \frac{6\sigma^{2}}{\pi^{2} A^{2}N^{2}(N^{2}-1)} \nonumber \\
\text{var}(\hat{f_{1}}) &\geq \frac{6\sigma^{2}}{\pi^{2} A^{2}N^{2}(N^{2}-1)} \nonumber
\end{empheq}
The CRLB of the amplitude and offset terms depend on known values, i.e. the dimension of image sub-block and the variance of the noise, while that of the frequencies and phase depend on an unknown parameter, i.e. the amplitude. 

\subsection{Maximum Likelihood Estimation (MLE) of Sinusoidal Parameters}

Recall that the log-likelihood function of $\boldsymbol{\theta}$ is 

\begin{align*}
\ln p({\bf s};\boldsymbol{\theta})= {} & \ln(c)-\frac{1}{2\sigma^{2}}({\bf s}-{\bf z}(\boldsymbol{\theta}))^{T}({\bf s}-{\bf z}(\boldsymbol{\theta}))
\end{align*}
In order to maximize the likelihood, we need to minimize the squared error:

\begin{equation*}
J(\boldsymbol{\theta})=({\bf s}-{\bf z}(\boldsymbol{\theta}))^{T}({\bf s}-{\bf z}(\boldsymbol{\theta}))
\end{equation*}
The estimator $\boldsymbol{\hat{\theta}}$ that minimizes the squared error $J$ is the maximum likelihood estimator.

We will return to the squared error, but let's rewrite Eq. (\ref{eq:model}) as 

\footnotesize
\begin{equation*}
f(x,y)=A\cos\phi\sin[2\pi(f_{0}x+f_{1}y)] + A\sin\phi\cos[2\pi(f_{0}x+f_{1}y)] + B,
\end{equation*}
\normalsize
where $x=0,...,N-1; \; y=0,...N-1$. Let ${\bf u}$ and ${\bf v}$ be the $N^{2} \times 1$ vectors, respectively, denoting the array of $\sin$ and $\cos$ terms in Eq. (\ref{eq:model}). Denote $\alpha_{1}=A\cos\phi$, $\alpha_{2}=A\sin\phi$, and $\boldsymbol{\alpha}=(\alpha_{1} \; \alpha_{2} \; B)$.  Further let $H=[{\bf u} \; {\bf v} \; {\bf 1}]$, which is $N^{2} \times 3$. Now we can rewrite the squared error as

%\scriptsize
%\begin{align*}
%{\bf u}= {} & [\sin(0) \; \sin(2\pi f_{0}) \; \sin(4\pi f_{0}) \; \cdots \sin(2\pi(N-1)f_{0}) \; \sin(2\pi f_{1}) \\ & \sin(2\pi(f_{0}+f_{1})) \; \sin(4\pi f_{0}+2\pi f_{1}) \; \cdots \sin(2\pi(N-1)f_{0}+2\pi f_{1}) \\ & \sin(4\pi f_{1}) \; \cdots \sin(2\pi(N-1)(f_{0}+f_{1}))]^{T}
%\end{align*}
%\normalsize
%and
%
%\scriptsize 
%\begin{align*}
%{\bf v}= {} & [\cos(0) \; \cos(2\pi f_{0}) \; \cos(4\pi f_{0}) \; \cdots \cos(2\pi(N-1)f_{0}) \; \cos(2\pi f_{1}) \\ & \cos(2\pi(f_{0}+f_{1})) \; \cos(4\pi f_{0}+2\pi f_{1}) \; \cdots \cos(2\pi(N-1)f_{0}+2\pi f_{1}) \\ & \cos(4\pi f_{1}) \; \cdots \cos(2\pi(N-1)(f_{0}+f_{1}))]^{T}
%\end{align*}
%\normalsize
%Both vectors are each $N^{2} \times 1$. Denote $\alpha_{1}=A\cos\phi$, $\alpha_{2}=A\sin\phi$, and $\boldsymbol{\alpha}=(\alpha_{1} \; \alpha_{2} \; B)$.  Further let $H=[{\bf u} \; {\bf v} \; {\bf 1}]$, which is %$N^{2} \times 3$. 
%Recall that the $N^{2} \times 1$ vector of signal measurements is
%
%\begin{align*}
%{\bf s}= {} & [s(0,0) \; s(1,0) \; s(2,0) \; \cdots s(N-1,0) \; s(0,1) \; s(1,1) \; \\ & s(2,1) \; \cdots s(N-1,1) \; s(0,2) \; \cdots s(N-1,N-1)]^{T}
%\end{align*}
%Now we can rewrite the squared error as

%\begin{align*}
%J(\boldsymbol{\alpha},f_{0},f_{1}) = {} & ({\bf s}-{\bf z})^{T}({\bf s}-{\bf z}) \\ = {} & ({\bf s}-\alpha_{1}{\bf u}-\alpha_{2}{\bf v}-{\bf B})^{T}({\bf s}-\alpha_{1}{\bf u}-\alpha_{2}{\bf v}-{\bf B}) \\
%= {} & ({\bf s}-{\bf H}\boldsymbol{\alpha})^{T}({\bf s}-{\bf H}\boldsymbol{\alpha})
%\end{align*}
\begin{align*}
J(\boldsymbol{\alpha},f_{0},f_{1}) = {} & ({\bf s}-\alpha_{1}{\bf u}-\alpha_{2}{\bf v}-{\bf B})^{T}({\bf s}-\alpha_{1}{\bf u}-\alpha_{2}{\bf v}-{\bf B}) \\
= {} & ({\bf s}-{\bf H}\boldsymbol{\alpha})^{T}({\bf s}-{\bf H}\boldsymbol{\alpha})
\end{align*}

Optimizing $J$ with respect to $\boldsymbol{\alpha}$ yields

\begin{equation}
\label{eq:MLE}
\hat{\boldsymbol{\alpha}}=({\bf H}^{T}{\bf H})^{-1}{\bf H}^{T}{\bf s}
\end{equation}
so that 

%\small
%\begin{align*}
%J(\boldsymbol{\hat{\alpha}},f_{0},f_{1})= {} & ({\bf s}-{\bf H}\boldsymbol{\hat{\alpha}})^{T}({\bf s}-{\bf H}\boldsymbol{\hat{\alpha}}) \\ = {} & ({\bf s}-{\bf H}({\bf H}^{T}{\bf H})^{-1}{\bf H}^{T}{\bf s})^{T}({\bf s}-{\bf H}({\bf H}^{T}{\bf H})^{-1}{\bf H}^{T}{\bf s}) \\ = {} & {\bf s}^{T}({\bf I}_{N^{2} \times N^{2}}-{\bf H}({\bf H}^{T}{\bf H})^{-1}{\bf H}^{T}){\bf s}
%\end{align*}
%\normalsize 
\small
\begin{align*}
J(\boldsymbol{\hat{\alpha}},f_{0},f_{1})= {} & ({\bf s}-{\bf H}({\bf H}^{T}{\bf H})^{-1}{\bf H}^{T}{\bf s})^{T}({\bf s}-{\bf H}({\bf H}^{T}{\bf H})^{-1}{\bf H}^{T}{\bf s}) \\ = {} & {\bf s}^{T}({\bf I}_{N^{2} \times N^{2}}-{\bf H}({\bf H}^{T}{\bf H})^{-1}{\bf H}^{T}){\bf s}
\end{align*}
\normalsize 
Minimizing $J$ is now equivalent to maximizing ${\bf s}^{T}{\bf H}({\bf H}^{T}{\bf H})^{-1}{\bf H}^{T}{\bf s}$, or equivalently,

%\begin{align}
%\label{eq:error}
%{\bf s}^{T}{\bf H}({\bf H}^{T}{\bf H})^{-1}{\bf H}^{T}{\bf s},
%\end{align}
%or equivalently,

$$
\begin{pmatrix}
{\bf s}^{T}{\bf u} & {\bf s}^{T}{\bf v} & {\bf s}^{T}{\bf 1} 
\end{pmatrix}
\begin{pmatrix}
{\bf u}^{T}{\bf u} & {\bf u}^{T}{\bf v} & {\bf u}^{T}{\bf 1} \\
{\bf v}^{T}{\bf u} & {\bf v}^{T}{\bf v} & {\bf v}^{T}{\bf 1} \\
{\bf 1}^{T}{\bf u} & {\bf 1}^{T}{\bf v} & {\bf 1}^{T}{\bf 1} 
\end{pmatrix}^{\!\!-1}
\begin{pmatrix}
{\bf u}^{T}{\bf s} \\
{\bf v}^{T}{\bf s} \\
{\bf 1}^{T}{\bf s}
\end{pmatrix}
$$
Noting that

%\begin{align*}
\begin{equation*}
\begin{split}
{\bf u}^{T}{\bf u}=\sum_{x=0}^{N-1}\sum_{y=0}^{N-1}\sin^{2}[2\pi(f_{0}x+f_{1}y)] \approx \frac{N^{2}}{2}  \\
{\bf u}^{T}{\bf v}=\sum_{x=0}^{N-1}\sum_{y=0}^{N-1}\sin[2\pi(f_{0}x+f_{1}y)]\cos[2\pi(f_{0}x+f_{1}y)] \approx 0 \\
{\bf u}^{T}{\bf 1}=\sum_{x=0}^{N-1}\sum_{y=0}^{N-1}\sin[2\pi(f_{0}x+f_{1}y)] \approx 0 \\
{\bf u}^{T}{\bf u}=\sum_{x=0}^{N-1}\sum_{y=0}^{N-1}\cos^{2}[2\pi(f_{0}x+f_{1}y)] \approx \frac{N^{2}}{2}  \\
{\bf u}^{T}{\bf u}=\sum_{x=0}^{N-1}\sum_{y=0}^{N-1}\cos[2\pi(f_{0}x+f_{1}y)] \approx 0
\end{split}
\end{equation*}
%\end{align*}
and simplifying yields

%\scriptsize
%$$
%\begin{pmatrix}
%\sum_{x}\sum_{y}s(x,y)\sin[2\pi(f_{0}x+f_{1}y)] \\
%\sum_{x}\sum_{y}s(x,y)\cos[2\pi(f_{0}x+f_{1}y)]  \\
%\sum_{x}\sum_{y}s(x,y)
%\end{pmatrix}^{\!\! T}
%\begin{pmatrix}
%\frac{N^{2}}{2} & 0 & 0 \\
%0 & \frac{N^{2}}{2} & 0 \\
%0 & 0 & N^{2} 
%\end{pmatrix}^{\!\!-1}
%\begin{pmatrix}
%\sum_{x}\sum_{y}s(x,y)\sin[2\pi(f_{0}x+f_{1}y)] \\
%\sum_{x}\sum_{y}s(x,y)\cos[2\pi(f_{0}x+f_{1}y)] \\
%\sum_{x}\sum_{y}s(x,y)
%\end{pmatrix}
%$$
%\normalsize
%Simplifying, we obtain 

\footnotesize
\begin{align*}
{\bf s}^{T}{\bf H}({\bf H}^{T}{\bf H})^{-1}{\bf H}^{T}{\bf s} \approx {} & \frac{2}{N^{2}}\left(\sum_{x}\sum_{y}s(x,y)\sin[2\pi(f_{0}x+f_{1}y)] \right)^{2}  \\ &  + \frac{2}{N^{2}}\left(\sum_{x}\sum_{y}s(x,y)\cos[2\pi(f_{0}x+f_{1}y)] \right)^{2}  \\ & +  \frac{1}{N^{2}}\left(\sum_{x}\sum_{y}s(x,y) \right)^{2}
\end{align*}
\normalsize
Recall that the Fourier transform (FT) of a function $f(x,y)$ is $F(f_{x},f_{y})=\sum_{x}\sum_{y}f(x,y)e^{-2\pi i(f_{x}x+f_{y}y)}$. Denoting the FT of $s(x,y)$ as $S(f_{0},f_{1})$, we lastly obtain

\small
\begin{equation}
\label{eq:error2}
{\bf s}^{T}{\bf H}({\bf H}^{T}{\bf H})^{-1}{\bf H}^{T}{\bf s} \approx \frac{2}{N^{2}} | S(f_{0},f_{1}) |^{2} + \frac{1}{N^{2}}\left(\sum_{x}\sum_{y}s(x,y) \right)^{2},
\end{equation}
\normalsize
where $| S(f_{0},f_{1}) |^{2}$ denotes the periodogram of $s(x,y)$. Since the second term in Eq. (\ref{eq:error2}) is fixed, the expression ${\bf s}^{T}{\bf H}({\bf H}^{T}{\bf H})^{-1}{\bf H}^{T}{\bf s}$ is maximized when the periodogram of the signal is maximized. 

The frequencies at which the periodogram is maximized have to be found numerically. Denote the optimal frequencies as $\hat{f_{0}}$ and $\hat{f_{1}}$; now Eq. (\ref{eq:MLE}) becomes

$$
\begin{pmatrix}
\hat{A}\cos\hat{\phi} \\
\hat{A}\sin\hat{\phi} \\
\hat{B}
\end{pmatrix}
=
\begin{pmatrix}
\frac{2}{N^{2}}\sum_{x}\sum_{y}s(x,y)\sin[2\pi(\hat{f_{0}}x+\hat{f_{1}}y)] \\
\frac{2}{N^{2}}\sum_{x}\sum_{y}s(x,y)\cos[2\pi(\hat{f_{0}}x+\hat{f_{1}}y)] \\
\frac{1}{N^{2}}\sum_{x}\sum_{y}s(x,y)
\end{pmatrix}
$$

Hence, the maximum likelihood estimators of $\boldsymbol{\hat{\theta}}$ in Eq. (\ref{eq:model}) are

%\begin{equation}
%\label{eq:final}
%\begin{split}
%(\hat{f_{0}},\hat{f_{1}}) = {} & \max_{f_{0},f_{1}}|F(f_{0},f_{1})|^{2} \nonumber \\
%\hat{A} = {} & \frac{2}{N^{2}} | S(\hat{f_{0}},\hat{f_{1}}) | \nonumber \\ %\sqrt{ \left( \sum_{x}\sum_{y}s(x,y)\sin[2\pi(\hat{f_{0}}x+\hat{f_{1}}y)] \right)^{2} + \left( \sum_{x}\sum_{y}s(x,y)\cos[2\pi(\hat{f_{0}}x+\hat{f_{1}}y)] \right)^{2}}  \\
%\hat{B} = {} & \frac{1}{N^{2}}\sum_{x}\sum_{y}s(x,y)  \nonumber \\
%\hat{\phi} = {} & \arctan \left( \frac{\sum_{x}\sum_{y}s(x,y)\cos[2\pi(\hat{f_{0}}x+\hat{f_{1}}y)]}{\sum_{x}\sum_{y}s(x,y)\sin[2\pi(\hat{f_{0}}x+\hat{f_{1}}y)]} \right) \nonumber
%\end{split}
%\end{equation}
\begin{empheq}[box=\fbox]{align}
\label{eq:final}
(\hat{f_{0}},\hat{f_{1}}) &= \max_{f_{0},f_{1}}|F(f_{0},f_{1})|^{2} \nonumber \\
\hat{A} &= \frac{2}{N^{2}} | S(\hat{f_{0}},\hat{f_{1}}) | \nonumber \\ 
\hat{B} &= \frac{1}{N^{2}}\sum_{x}\sum_{y}s(x,y)  \nonumber \\
\hat{\phi} &= \arctan \left( \frac{\sum_{x}\sum_{y}s(x,y)\cos[2\pi(\hat{f_{0}}x+\hat{f_{1}}y)]}{\sum_{x}\sum_{y}s(x,y)\sin[2\pi(\hat{f_{0}}x+\hat{f_{1}}y)]} \right) \nonumber
\end{empheq}
Here, $F(f_{0},f_{1})=\sum_{x}\sum_{y}f(x,y)e^{-2\pi i(f_{0}x+f_{1}y)}$, i.e. the 2D discrete Fourier transform (FT) of Eq. (\ref{eq:model}), and $| F(f_{0},f_{1}) |^{2}$ denotes the periodogram of $f(x,y)$. The frequencies at which the periodogram is maximized, $(\hat{f_{0}},\hat{f_{1}})$, have to be found numerically. Note that the maximum likelihood estimator of the offset term $B$ is simply the mean of the signal measurements, while the maximum likelihood estimator of the amplitude is the magnitude of the FT of the signal evaluated at the optimal frequencies. 

\section{Size of $N$}

\begin{figure}
\begin{center}

\subfloat[Phase $\phi=0$]{\includegraphics[width=1.5in]{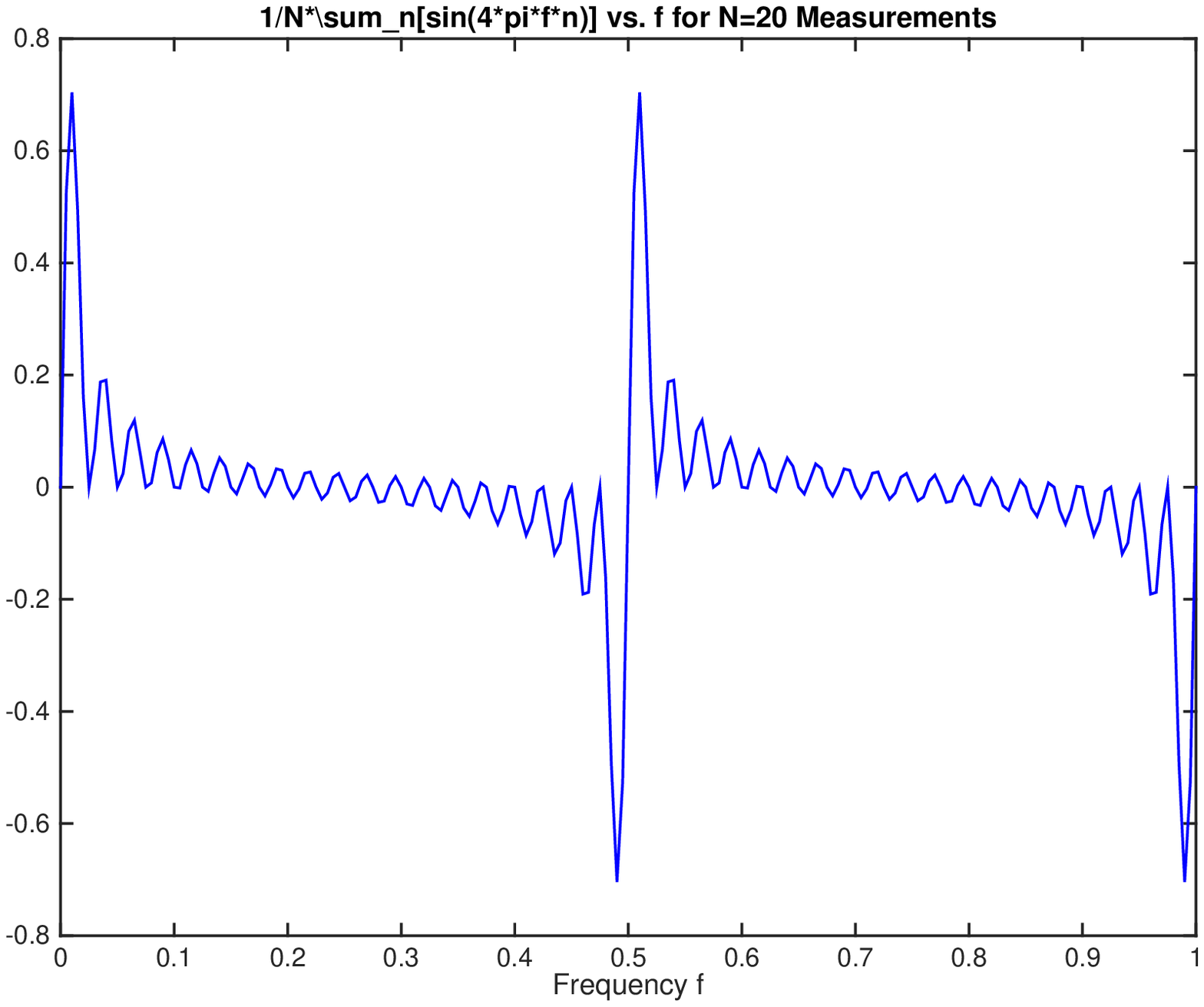}}\quad
\subfloat[Phase $\phi=\pi/4$]{\includegraphics[width=1.5in]{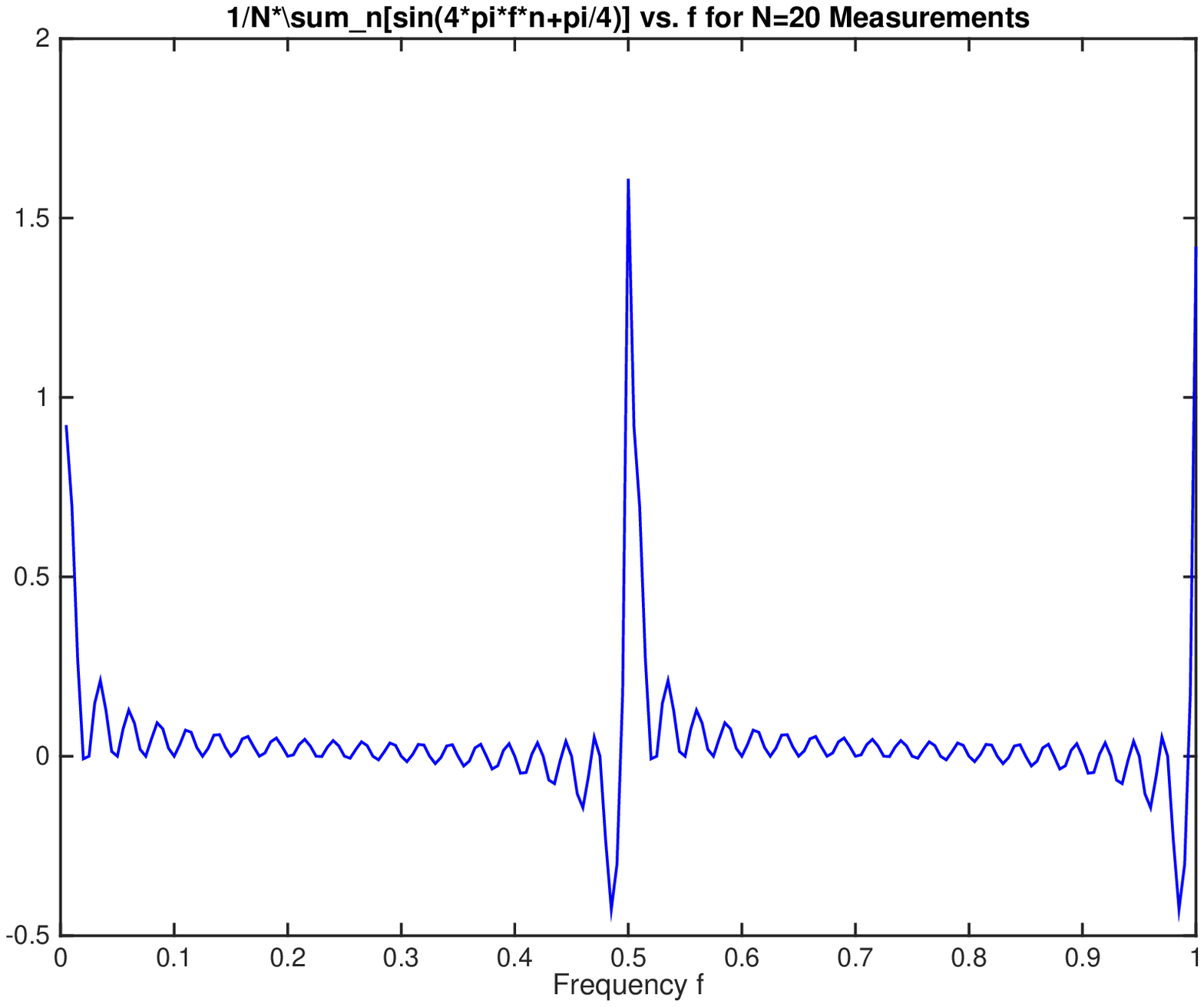}}
 
\caption{Plot of $y(f)=\frac{1}{N}\sum_{x=0}^{N-1}\sin(4\pi fx + \phi)$ for $N=20$ samples for two different phases. In both cases, if $f$ is not near $0$, $1/2$, or $1$, then $y(f)$ is approximately zero. As $N$ increases, $y(f)$ becomes closer to zero for $f$ not near $0$, $1/2$, or $1$.}
\label{fig:f1}

\end{center}  
\end{figure}

\begin{figure}
\begin{center}

\subfloat[Phase $\phi=0$]{\includegraphics[width=1.5in]{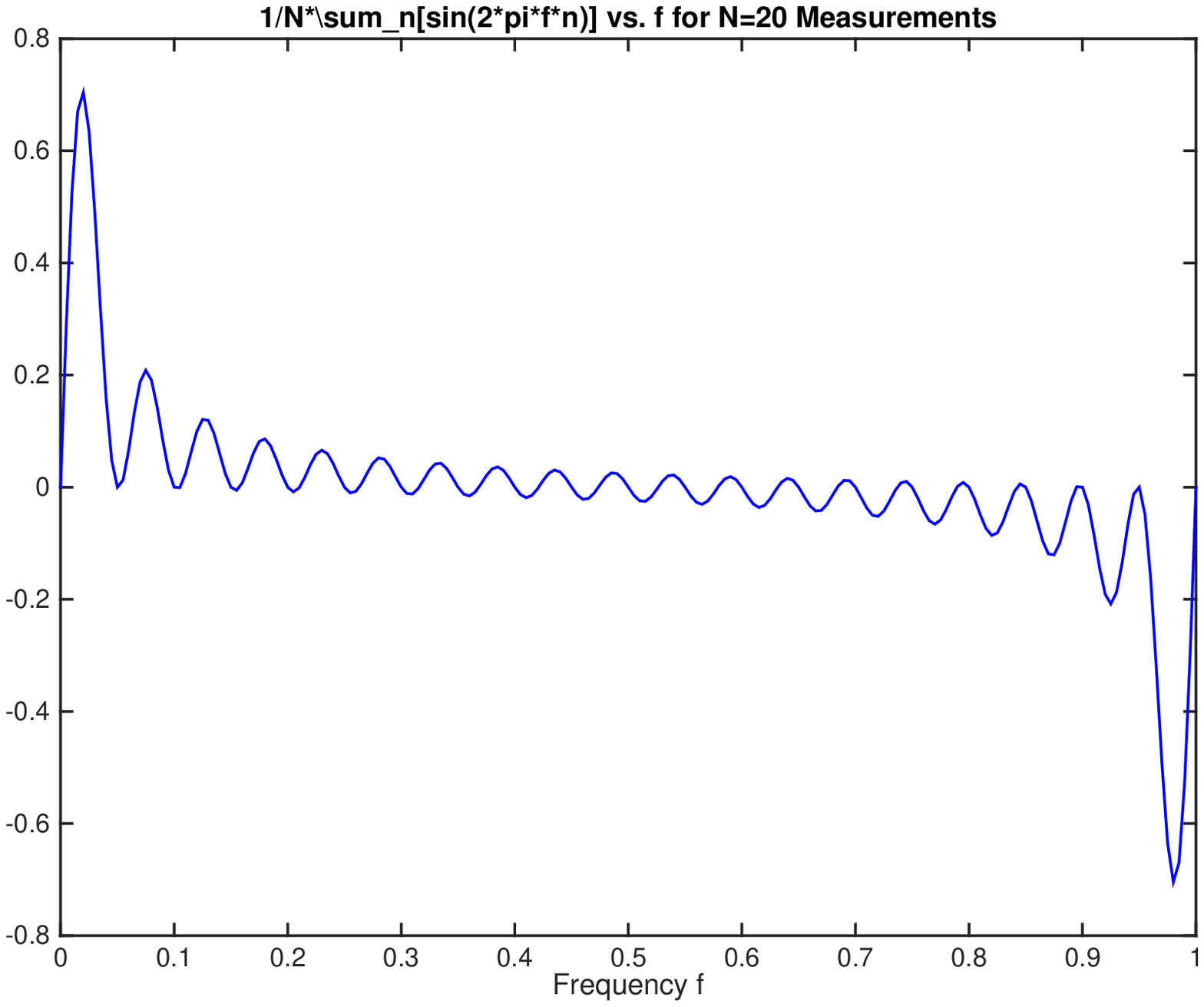}}\quad
\subfloat[Phase $\phi=\pi/4$]{\includegraphics[width=1.5in]{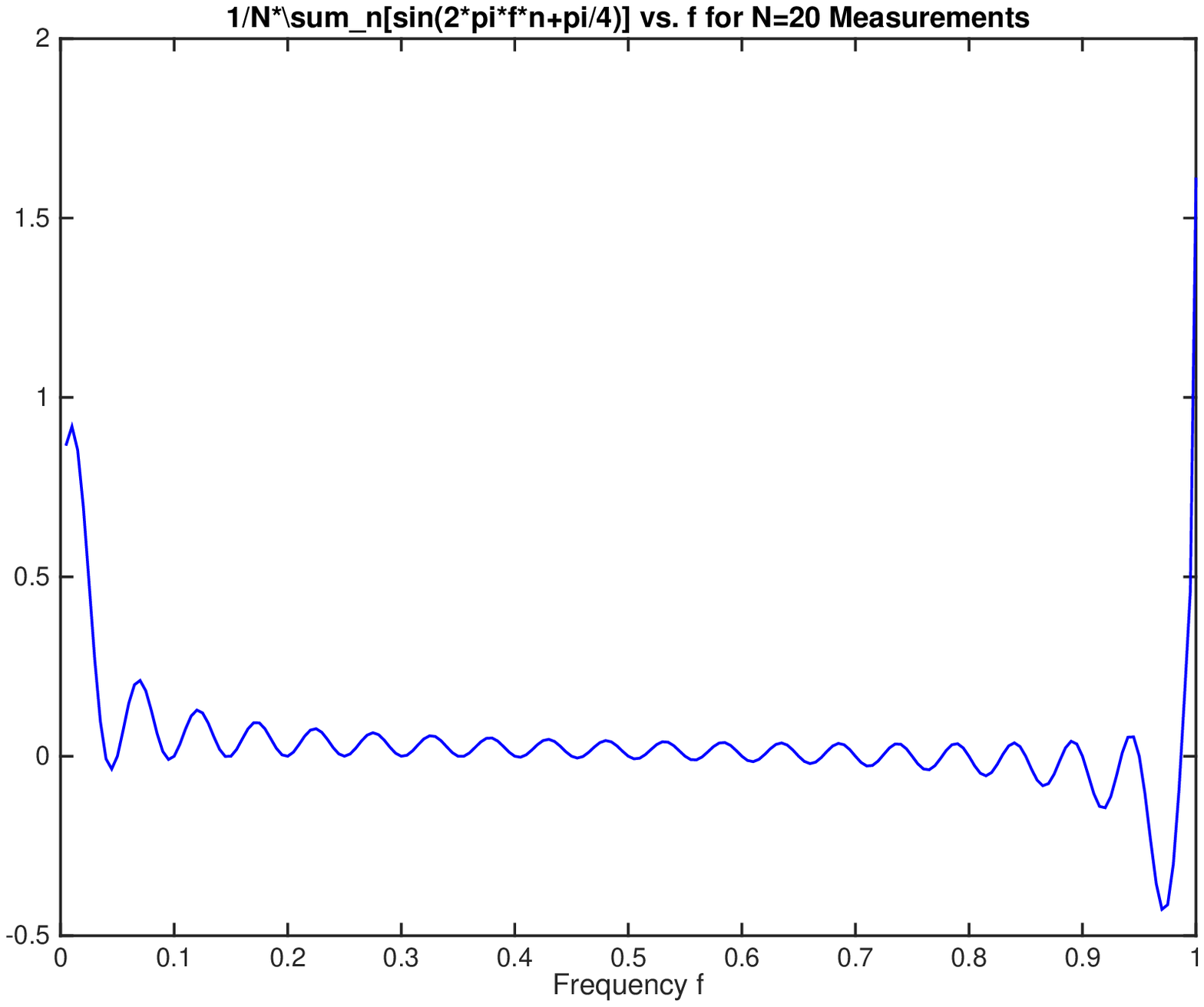}}
 
\caption{Plot of $y(f)=\frac{1}{N}\sum_{x=0}^{N-1}\sin(2\pi fx + \phi)$ for $N=20$ samples for two different phases. In both cases, if $f$ is not near $0$ or $1$, then $y(f)$ is approximately zero. As $N$ increases, $y(f)$ becomes closer to zero for $f$ not near $0$ or $1$.}
\label{fig:f2}

\end{center}  
\end{figure}

To get an idea of how big the dimension $N$ should be in order for the approximation
\begin{equation}
\label{eq:app}
\frac{1}{N}\sum_{x}e^{i(2k\pi f \; x+\phi)} \approx 0, \; \; k=1,2
\end{equation}
to hold, we look at the plots of two functions: 1) $y(f)=\frac{1}{N}\sum_{x=0}^{N-1}\sin(4\pi fx + \phi)$ and 2) $y(f)=\frac{1}{N}\sum_{x=0}^{N-1}\sin(2\pi fx + \phi)$ for $\phi=0,\pi/4$ and $N=20$ measurements. These plots are shown in Figs. \ref{fig:f1} and \ref{fig:f2}.

For the $\omega=4\pi f$ case, shown in Fig. \ref{fig:f1}, if $f$ is not near 0, 0.5, or 1, the summation is approximately zero. For the $\omega=2\pi f$ case, shown in Fig. \ref{fig:f2}, if $f$ is not near 0 or 0.5, the summation is approximately zero; however, it has a slower approximation to zero than the $\omega=4\pi$ case. The plots illustrate that $N=20$ measurements is adequate for Eq. (\ref{eq:app}) to be valid. 
% the derived (approximate) CRLB and MLE solutions are valid for $0 \leq f \geq 0.5$. 

%\section{Acknowledgements}
%The authors would like to acknowledge insightful discussions with Dr. Alex Ushveridze. 

\bibliographystyle{IEEEtran}
\bibliography{refs}

\begin{thebibliography}{1}
\providecommand{\url}[1]{#1}
\csname url@samestyle\endcsname
\providecommand{\newblock}{\relax}
\providecommand{\bibinfo}[2]{#2}
\providecommand{\BIBentrySTDinterwordspacing}{\spaceskip=0pt\relax}
\providecommand{\BIBentryALTinterwordstretchfactor}{4}
\providecommand{\BIBentryALTinterwordspacing}{\spaceskip=\fontdimen2\font plus
\BIBentryALTinterwordstretchfactor\fontdimen3\font minus
  \fontdimen4\font\relax}
\providecommand{\BIBforeignlanguage}[2]{{%
\expandafter\ifx\csname l@#1\endcsname\relax
\typeout{** WARNING: IEEEtran.bst: No hyphenation pattern has been}%
\typeout{** loaded for the language `#1'. Using the pattern for}%
\typeout{** the default language instead.}%
\else
\language=\csname l@#1\endcsname
\fi
#2}}
\providecommand{\BIBdecl}{\relax}
\BIBdecl

\bibitem{rife.1974}
D.~Rife and R.~Boorstyn, ``Single-tone parameter estimation from discrete-time
  observations,'' \emph{IEEE Trans. Information Theory}, vol.~20, pp. 591--598,
  1974.

\bibitem{rife.1976}
D.~Rife and R.~Boorsten, ``Multiple tone parameter estimation from
  discrete-time observations,'' \emph{Bell System Technical Journal}, pp.
  1389--1410, 1976.

\bibitem{lang.1980}
S.~Lang and J.~McClellan, ``Frequency estimation with maximum entropy spectral
  estimators,'' \emph{IEEE Trans. Pattern Acoustics, Speech, and Signal
  Processing}, vol.~28, pp. 716--724, 1980.

\bibitem{stoica.1989}
P.~Stoica, R.~Moses, B.~Friedlander, and T.~Soderstrom, ``Maximum likelihood
  estimation of the parameters of multiple sinusoids from noisy measurements,''
  \emph{IEEE Trans. Acoustics, Speech, and Signal Processing}, vol.~37, pp.
  378--392, 1989.

\bibitem{hainsworth.2003}
S.~Hainsworth and M.~Macleod, ``On sinusoidal parameter estimation,'' in
  \emph{Proc. of the 6$^{th}$ Int. Conference on Digital Audio Effects}, 2003.

\end{thebibliography}


%\end{IEEEbiography}


%\newpage



%\vfill


\end{document}